\documentclass[preprint,showpacs,preprintnumbers,amsmath,amssymb,nofootinbib]{revtex4}
\usepackage{graphicx}
\usepackage{dcolumn}
\usepackage{bm}

\begin{document}
\title{\bf Field Theories Found Geometrically from Embeddings in Flat Frame Bundles}
\author{Frank B. Estabrook}
\email{frank.b.estabrook@jpl.nasa.gov}
\affiliation{Jet Propulsion Laboratory, California Institute of Technology, Pasadena, CA 91109}

\date{\today}

\begin{abstract}
  
  We present two families of exterior differential systems (EDS) for
  non-isometric embeddings of orthonormal frame bundles over
  Riemannian spaces of dimension $q = 2, 3, 4, 5....$ into orthonormal
  frame bundles over flat spaces of sufficiently higher dimension.  We have calculated Cartan characters showing that these EDS satisfy Cartan's
  test and are well-posed dynamical field theories.  The first family
  includes a constant-coefficient (cc) EDS for classical Einstein vacuum
  relativity ($q = 4$).  The second family is generated only by cc 2-forms, so these are integrable (but nonlinear) systems of partial
  differential equations. These latter field theories apparently are
  new, although the simplest case $q = 2$ turns out to embed a ruled
  surface of signature (1,1) in flat space of signature (2,1). Cartan
  forms are found to give explicit variational principles for all
  these dynamical theories.

\end{abstract}
\maketitle
\baselineskip= 14pt

\section{Introduction}

We discuss two families of geometric field theories.  By ``geometric"
we mean that these theories are given as exterior differential systems
(EDS) for embedding of q-dimensional submanifolds ${R^q}$ in flat
homogeneous isotropic metric spaces \({E^N}\) of higher dimension, say
$N$. To formulate these EDS we in fact embed the orthonormal frame
bundles over the submanifolds into the orthonormal frame bundles over
the flat spaces, that is, into the groups ISO($N$), which have
dimension $N(N + 1)/2$. E. g., the $q = 4$ dimensional EDS are set
using the 55 basis 1-forms of ISO(10).  The fibers of the embedded
bundles are subgroups of the O($N$) fibers of ISO($N$), thus inducing
embedding maps of their q-dimensional bases ${R^q}$ into the ${E^N}$
bases of ISO($N$).
        
By ``field theories'' we mean that each of these various EDS is shown,
by an explicit numerical calculation of its Cartan characteristic
integers \cite{Ivey} to have the property of being well-posed or, with
the correct signature, of being``causal''.  This calculation uses a
suite of {\itshape Mathematica} programs for EDS written by H. D.
Wahlquist, and evaluates the Cartan characteristic integers with a
Monte Carlo program to compute the ranks of the large matrices that
arise \cite{Wahlquist}. The successive integers determine the
dimension and well-posedness of the general solutions, and the
Wahlquist programs also confirm the ``involutory'' property of certain fields in a solution, viz. those that can be adopted as independent variables.
        
A properly set EDS in a space with N variables is equivalent to a set
of first order partial differential equations in N-q dependent
variables, functions of q independent variables, and Cartan's
technique of EDS is a deep approach to the Cauchy-Kowalewska analysis
of such field theories.
        
Cartan's theory considers construction of a sequence of regular integral manifolds (of
successively higher dimensions) of an EDS. His characteristic integers $s_{i}, i=0,1,..,q-1$,
are calculated from the ranks of matrices that arise, and are
diagnostic; they must pass Cartan's test \cite{Ivey} if the EDS (or
an equivalent set of partial differential equations) is well posed.
Then the final construction of the solution is determined solely from gauge and 
boundary data, and, at least in the analytic category, Cauchy
existence and uniqueness are proved.  We believe that, with proper attention to signature, the sets of partial differential equations
following from such a well-posed EDS are those of a canonical field
theory.  Non-trivial embedding EDS that are well-posed, or causal, are not common.  A key to their existence may be that for all the EDS we consider here we are also able to find Cartan $q$-forms from which the EDS may be derived by arbitrary variation.

There is a large literature, beginning with Lepage and Dedecker, on
the use of Cartan $q$-forms and their closure $q + 1$-forms
(``multisymplectic'' forms).  These are respectively the
multidimensional field-theoretic extensions of classical Hamiltonian
theory (the Cartan 1-form $L dt$) and symplectic geometry.  A short
but essential bibliography can be found in Gotay \cite{got}; cf. also
Gotay et al \cite{GIMMsy}, Bryant et al \cite{Bryant}, Hermann \cite{her88} and Estabrook
\cite{est80} \cite{est06}.  The differential geometric setting for
that work was for the most part (the structure equations of basis
forms on) the first or second jet bundle over a base of q independent
variables.  Our use instead of basis forms and structure equations for
embedding geometry, orthonormal frame bundles over flat metric
geometries, is more in the spirit of string theory.  It allows application of the
variational techniques of field theory to the movable frames of
general relativity, and can lead to interesting extensions.

In both these families of EDS we adapt the method usually used in
the mathematical literature for isometric embedding, cf. e.g.
\cite{bry83}, \cite{gri87}, in that we do not begin with a prior framing of the solution and prolong to higher bundles, but rather only use the bases of the embedding bundle.  Such EDS have also been used in the theory of calibrated subspaces \cite{Ivey}.  Perhaps this generalization of
the customary isometric embedding can be called ``dynamic embedding''.  The EDS that naturally arise are considerably more elegant, interpretable as field theories.

The Lie group ISO($N$) (or one of its signature siblings ISO($N-1, 1$)
etc.) is the isometry group of $N$-dimensional flat space ${E^N}$ (or a signature sibling).
The group space is spanned by $P=N(N + 1)/2$ canonical vector fields,
and by a dual basis of left-invariant 1-forms that we first denote by
$\theta ^\mu , \mu = 1...N$, corresponding to translations, and
${\omega ^\mu}_ \nu $, that will correspond to rotations.  Now the
structure equations for general movable frames over an $N$-dimensional
manifold are usually written covariantly (on the second frame bundle)
as
\begin{eqnarray}
{d\theta }^{\mu}+{\omega^{\mu}}_{ \nu} \wedge {\theta }^{\nu } &=& 0 \\
{d\omega^{\mu }}_{\nu }+{\omega ^{\mu }}_{\sigma }\wedge {\omega
^{\sigma }}_{\nu }+{R^{\mu }}_{\nu
} &=& 0.
\end{eqnarray}

These become the Cartan-Maurer equations of ISO($N$) or one of its
siblings when the curvature 2-forms \({{{R^{\mu }}}_{\nu }}\) are put
equal to zero, and upper indices are systematically lowered using (for
signature) a non-singular matrix of constants \({{\eta }_{{\mu \nu}}}\), after which imposing antisymmetry (orthonormality)
${{\omega }_{{\mu \nu }}} = - {{\omega }_{{\nu \mu }}}$ .  These
structure equations then describe $N(N-1)/2$-dimensional rotation
groups as fibers over $N$-dimensional homogeneous spaces $E^N$.  (The $\eta_{\mu \nu}$, and other possible signatures in $E^N$, are often conveniently ignored in the following, and can be inserted later.) 

We will write the two families of EDS using partitions $(n, m), n + m
= N$, of the basis forms of ISO($N$) into classes labeled respectively
by the first $n$ indices $i, j$, etc. $= 1, 2, ... n$ and the
remaining indices $A, B$, etc. $= n +1, n +2, ... N$.  So the basis
forms are ${{\theta }^i}, {{\theta }^A}$, and, after lowering an
index, ${{\omega }_{{ij}}} = - {{\omega }_{{ji}}}$, ${{\omega }_{{AB}}}
= - {{\omega }_{{BA}}}$, ${{\omega }_{{iA}}} =-{{\omega }_{{Ai}}}$.
Summation conventions on repeated indices will be used separately on
each partition.  The structure equations (1) (2) before lowering become
\begin{eqnarray}
d{\theta ^i}+{\omega ^i}_j \wedge {\theta ^j} &=& -{\omega ^i}_A
\wedge {\theta ^A} \\
{d\theta ^A}+{\omega ^A}_B \wedge {\theta ^B} &=& -{\omega ^A}_i
\wedge {\theta ^i} \\
{d\omega ^i}_j+{\omega ^i}_k \wedge {\omega ^k}_j &=& - {\omega ^i}_A
\wedge
{\omega ^A}_j \\
{d\omega ^A}_B +{\omega ^A}_C \wedge {\omega ^C}_B &=&-{\omega ^A}_i
\wedge
{\omega ^i}_B  \\
{d\omega ^i}_A +{\omega ^i}_j \wedge {\omega ^j}_A + 
{\omega ^i}_B \wedge {\omega ^B}_A &=& 0.
\end{eqnarray}
The terms we have put on the right are interpreted as torsions and
curvatures induced by an embedding; we will use them to set the EDS.

In Sections 2 and 3 we calculate the Cartan characteristic integers of
the embedding EDS for the two families.  We will report the results in
a short tabular form: $P\{s_0, s_1,...s_{q-1}\}q+CC$.  This gives
first the dimension $P$ of the space in which we set the EDS, i.e. the
total number of basis forms with whose structure equations we begin,
then the series of Cartan integers found, $\{s_0, s_1,...s_{q-1}\}$.  $q$ is the dimensionality of the base space of a general solution. Finally CC is
the number of Cartan characteristic vectors (the number of auxiliary
fields allowing us to write a cc system, fibers corresponding to
variables that could in principle be eliminated from the EDS).  Cartan
denotes $q + CC$ by $g$, the genus.  The ultimately simple Cartan test
showing the EDS to be well-posed and causal, calculated from these, is
derived in \cite{Ivey} and the literature cited there. Here the test
is simply that these integers satisfy $P-\sum_{i=0}^{q-1} s_i - q - CC := s_{q} \geq 0$.  We will always have $s_q =0$, which a physicist interprets as
absence of a gauge group, and so according to Cartan solutions will depend on $s_{q-1}$
functions of q-1 variables. We denote these theories as causal but
that requires also adjusting the signatures, so that the final
integration of solutions from this boundary data is hyperbolic.

It is a classic result \cite{gri87} that smooth local embedding of
Riemannian geometries of dimension $q = 3, 4, 5 ...$ is always
possible into flat spaces of dimension respectively $N = q(q+1)/2=6, 10, 15
...$ , which motivates the partitions of our first family, viz.  $(n,
m) = (3, 3), (4, 6), (5, 10), ...$ The causal EDS we give determine
submanifolds of ISO($N$) which are themselves $O(n) \otimes O(m)$
bundles fibered over $q = n$-dimensional base spaces, say ${R^q}$ and
induce maps of these into ${E^N}$. The $n$ ${{\theta }_i}$ remain
independent (``in involution'') when pulled back to a solution bundle,
satisfying the structure equations of an orthonormal basis in any
cross section, and Equations (5) and (7) express embedding relations
that go back to Gauss and Codazzi. The solution bundle metric is the
pullback of ${{\theta }_i} {{\theta }_i}$.  We will present in Section
2 the family of Einstein-Hilbert Cartan forms from which the EDS of
this first family are derived by variation.  The EDS will require zero
torsion for the \({{\theta }_i}\) but not insist on aligning the
solutions with these orthonormal frames (the \({{\theta }_A}\) are {\itshape not} included in the EDS so it is not necessarily ``isometric''), and
from the induced map of bundle bases there is also a less interesting
``ghost'' metric which is the pullback of ${{\theta }_i}{{\theta
  }_i}+{{\theta }_A}{{\theta }_A}$.  The induced curvature 2-forms are
required by the EDS to satisfy ``horizontality'' 3-form conditions and
also to have vanishing Ricci $n-1$-forms.

The field theories of our second family, of dimension $q = 2, 3, 4, 5
...$ also arise from embeddings into flat spaces ${E^N}$ of dimension
$N = 3, 6, 10, 15, ...$ but the EDS use different partitions, viz..,
$(n, m) = (1, 2), (2, 4), (3, 7), (4, 11)$, etc.  Solutions are ($O(n)
\otimes O(m)$ bundles over) geometries of dimension $q = n + 1$ and
can be called n-branes.  They have rulings that are flat n-spaces.
The EDS are generated only by cc sets of 2-forms (for vanishing torsion
of both partitions) and are so-called ``integrable systems''.  Again
the embedding is dynamic, the partitioned frames are not required to
be an orthonormal framing of the solution manifolds.  In Section 3 we give the EDS
and report the calculated Cartan characters showing them to be causal.  The
$n$ ${{\theta }_i}$ when pulled back into a solution both determine a Riemannian submersion and geodesic slicing.  This is either a theory of relativistic rigidity or perhaps
of a Kaluza-Klein gravitational field, depending on $N$ and the
signature adopted.  Cartan forms for those EDS are easily found.

As a sole illustration of introduction of explicit coordinates into such a frame bundle EDS, the
simplest of these non-isometric geometric field theories, that based
on partition $(1, 2)$, is integrated in Section 4.  Its solutions turn
out to be classically known, in the guise of geodesically ruled surfaces in
${E^3}$.  We have only changed signature to show it as a stringy field
causally evolving in time.

\section{Einstein-Hilbert Action}

The EDS of our first family arise from Cartan $n$-forms on ISO($N$)
expressing the Ricci scalars of $q = n$ dimensional submanifolds of
${E^N}$,
\begin{equation}
\Lambda  = {R_{{ij}}}\wedge {{\theta }_k}\wedge ...{{\theta }_p}
{{\epsilon }_{{ijk}...p}},
\end{equation}
where from the Gauss structure equation,  Eq. (5),  $2{R_{{ij}}} :=
-{{\omega }_{{iA}}}\wedge \)\({{\omega }_{{jA}}}$
is the induced Riemann 2-form.   The exterior derivative of the $n$-form
field $\Lambda $ on ISO($N$),  using Eq. (3) and (7),  is quickly
calculated to be the $n + 1$-form (closed,  multisymplectic)
\begin{equation}
{d\Lambda } = {{\theta }_A}\wedge {{\omega }_{{Ai}}}\wedge
{R_{{jk}}}\wedge {{\theta
}_l}\wedge ...{{\theta }_p} {{\epsilon }_{{ijkl}...p}}.
\end{equation}

This $n+1$-form is a sum of products of the $m$ 1-forms ${{\theta
  }_A}$ and the m $n$-forms ${{\omega }_{{Ai}}}\wedge \)\(
{R_{{jk}}}\wedge {{\theta }_l}\wedge \)...\({{\theta }_p}\)
\({{\epsilon }_{{ijkl}...p}}$.  A variational isometric embedding EDS
is generated by the m ${{\theta }_A}$, their exterior derivatives for
closure, and the m $n$-forms, since any vector field contracted on ${d\Lambda}$ yields a form in the EDS.  That is, up to
boundary terms, the arbitrary variation of \(\Lambda \) vanishes on
solutions.  We previously calculated Cartan's characteristic integers
for these isometric embedding EDS showing them to be well set and
causal \cite{est89} \cite{est99}.  We denoted them as being
``constraint free'' geometries.  Isometric embedding formulations of
the Ricci-flat field equations then are obtained by adding in the
closed $n-1$-forms for Ricci-flatness as constraints. The augmented
EDS are again calculated to be causal.  To be explicit, for partition (4, 6) the constraint-free Cartan character
table was 55 $\{6, 6, 6, 12\} q = 4 + 21$ which, with the
augumentation with four 3-forms became 55 $\{6, 6, 10, 8\} q = 4 +
21$.  We now see that formulation as nevertheless somewhat
unsatisfactory as field theory, since the Einstein-Hilbert action
appears to have lead to equations which in fact mostly follow from the
imposed constraints.

We have however noticed that there is another variational EDS
belonging to a different quadratic factoring of the multisymplectic forms
${d\Lambda }$, Eq. (9).  The ${{\theta }_{i }}$ will frame a
Riemannian metric on an embedded space of dimension $n$ so long as the
induced torsion 2-forms of Eq.(2), ${{\omega }_{{iA}}}\wedge {{\theta }_A}$,
vanish, and these factor Eq.(9) term-by-term, as products with the n
$n-1$-forms for Ricci-flatness. The exterior derivatives of the
torsion terms must be included;  these are sometimes called conditions for horizontality. In sum, we have
considered the following closed EDS (which now do {\itshape not}
include the mathematically customary isometric embedding 1-forms
${{\theta }_A}$)
\begin{equation}
({{\omega }_{{iA}}}\wedge {{\theta }_A},  {R_{{ij}}}\wedge {{\theta
}_j}, {R_{{ij}}}\wedge
{{\theta }_k}\wedge ...{{\theta }_l}{{\epsilon }_{{ijk}...{lp}}})
\end{equation}

When n=4 this EDS is an exact parallel to the EDS for a moving
frame formulation of vacuum relativity that used the 44 traditional intrinsic coordinates of
tetrad frames and connections over 4-space, and had 10 gauge freedoms \cite{est05}.  It had the same Cartan character table but no CC. The present formulation is
set with more variables, viz. 55, but its solutions have 21
CC fibers (since $\omega _{ij}$ and $\omega_{AB}$ do not
enter explicitly) and no gauge freedom; moreover it has the elegance of a
cc EDS (no coordinate functions appear in the
generating forms) \cite{est82} \cite{har83} \cite{har84}.

The calculation shows the EDS Eq.(10) to be well set and causal systems for
embedding of $O(n) \otimes O(m)$ bundles over $n$ space, for the
partitions (3, 3), (4, 6), (5, 10) etc. as stated in the introduction.
The embedding dimension, the computed Cartan characters, dimensionality and
$O(n) + O(m)$ fiber dimension (CC) of the solutions for these
cases are respectively $21\{0, 6, 3\}3 + 9, 55\{0, 4, 12, 14\}4 + 21, 120\{0, 5, 10, 20, 25\}5 + 55$, etc.  The base spaces of
the fibered solution manifolds are spanned by the 1-forms ${{\theta
  }_i}$; evidently a solution is a bundle of orthonormal frames
belonging to the Ricci-flat Riemannian connection ${{\omega
  }_{{ij}}}$.
The metric is ${{\theta }_i}{{\theta }_i}$. There is also present in
the base space ${R^n}$ another metric pulled back from the induced
embedding of it in the base space ${E^N}$ about which we know little: ${{\theta }_i}{{\theta }_i}+{{\theta }_A}{{\theta }_A}$ .
It is a ghost tensor field, perhaps with only indirect influence.  The
ideals we are writing are set on ISO($N$), and their solutions are
frame bundles embedded in ISO($N$), and the induced embeddings of the
base spaces seem to be of less interest.

The ideal Eq.(10) is contained in the augmented embedding
ideal we have previously used, so solutions of the latter will be
solutions of the former.  This would seem to imply that our new
dynamic embedding ideal will have additional solutions; indeed it
implies fewer partial differential equations than does the isometric
embedding ideal augmented with constraints for Ricci-flat geometry.
Perhaps so-called singular solutions of the isometric embedding
ideal--solutions which are not regular, that is, obtained by Cartan's
sequential integrations--now appear as regular solutions, which could
make this new formulation important for local numerical computation
from boundaries.

\section{Torsion-free  n-brane Embedding}

We have searched whether the torsion 2-forms induced in {\itshape
  both} the local partitions can together be taken as an EDS:
(${{\omega }_{{iA}}}\wedge {{\theta }_A}\), \({{\omega }_{{iA}}}\wedge
{{\theta }_i}$).
It can easily be checked that it is closed,  and calculation
of the characteristic integers indeed showed that
{\itshape for just the values of (n, m) of the second family described in the
introduction} these EDS are  causal,  with $q = n + 1$ and fibers $O(n) \otimes
O(m)$, dim $n(n-1)/2 + m(m-1)/2$.  The results for the first five EDS
are:
$(n, m) = (1, 2)$,  $6\{0, 3\}2 + 1$; $(2, 4)$, $21\{0, 6, 5\}
3 + 7$; $(3, 7)$, $55\{0, 10, 9, 8\}4$ + 24;  $(4, 11)$, $120\{0, 15, 14, 13,
12\} 5$ + 61;  $(5, 16)$, $231\{0, 21, 20, 19, 18, 17]6$ + 130; and the pattern
seems evident.

Now well set EDS for geodesic flat dimension n submanifolds of flat
N spaces are generated,  using the partition $(n, m)$,  by the closed ideal
of 1-forms $({{\theta }_{A }}, {{\omega }_{{Ai}}}$).
For example,  if $N = 3$ and $n = 1$ and $m = 2$,   geodesic lines in
flat 3-space,  the Cartan characteristic integers are
$6\{4\}1 +1$.  If $N = 4$,  for partition $(1, 3)$ we find $10\{6\}1 +
3$ (in all cases ${{\omega }_{{ij}}}$ and
${\omega }_{{AB}}$
give the Cauchy characteristic fibers).  Similarly,  the EDS for flat
2-dimensional submanifolds of flat $N$ spaces are generated by the 1-forms 
with partitions $(2, N-2)$.  For example if $N = 5$, $(n, m) = (2, 3)$,  and  the character table is $15\{9, 0\}2 + 4$. When $N = 6,  (n, m) = (2, 4)$ and $21\{12, 0\}2 + 7$.  The zeros can be ascribed to a gauge freedom.  These constructions clearly continue.  Our new torsion-free EDS
(${{\omega }_{{iA}}}\wedge {{\theta }_A}$,  ${{\omega }_{{Ai}}}\wedge
{{\theta }_i}$) are contained in (${{\theta }_{A }}, {{\omega
}_{{Ai}}}$), so we see that the q-dimensional solutions of the torsion-free
embedding theory must contain flat geodesic fibers of dimension $n = q-1$.  Thus 
the solutions are {\itshape ruled }spaces,

In a solution the \({{\theta }_{i }}\) remain independent (are ``in
involution'') but fall short by one of being a complete basis.  In addition to the slicing, they
define there a vector field, say {\itshape V}, of arbitrary
normalization (a congruence), by the relations \(V\cdot {{\theta
  }_i}\){\itshape } \(=\) \(V\cdot {{\omega }_{{ij}}}\) \(=\)
{\itshape }\(V\cdot {{\omega }_{{AB}}}\)\( \)\(=\) 0.  Contracting
{\itshape V} on the second torsion 2-form, since the ${{\theta }_i}$
remain linearly independent, gives also $V\cdot {\omega }_{iA}= 0$.
It follows that the Lie derivatives with respect to V of $\theta_{i}, \omega_{ij}$ and $R_{ij}$ vanish on solutions.  They live in (and are lifted from) an n dimensional quotient
space of the solution, with metric \({{\theta }_i}{{\theta }_i} \) and
Riemann tensor \({{\omega }_{{iA}}}\wedge {{\omega }_{{Aj}}} \).  Cross sections of this quotient map are the rulings, geodesic n-dimensional subspaces
calibrated by the volume form \({{\theta }_i}\wedge{{\theta
  }_j}...\wedge{{\theta }_k}\).

In an earlier time we have discussed the problem of defining a rigid
body in special and general relativity \cite{wah66}.  The kinematic
quotient-space definition of rigidity due originally to Max Born
(Riemannian submersion) was shown by Herglotz and Noether to have only
three degrees of freedom: the only Born-rigid congruences which were
rotating ( had vorticity) in Minkowski space were isometries of the
space-time without time evolution.  We showed this to be the case also
for kinematic or ``test'' rigid bodies moving in vacuum Einstein
spaces.  It seemed to be impossible then to sensibly discuss the
so-called ``dynamic'' rigid bodies envisioned by Pirani, which were to
carry their own 3-dimensional geometry while distorting space-time. We
are charmed by having now arrived at space-times, using dynamic
embedding in the (3, 7) partition, having the greater dynamical
freedom allowed by separation of the r\^oles of the induced 3-metrics
in the cross sections and quotient space of a solution.

In the (4, 11) partition, the solutions are five dimensional, with a
dynamically rigid congruence that projects to a metric quotient 4-space.  This
may be a well-posed causal variant of Kaluza-Klein theory, and
merits further investigation.

Closed EDS generated only by cc 2-forms have a special structure, inasmuch as they can be equivalent
to dual infinite Lie algebras of Ka\^c-Moody type and lead to
hierarchies of so-called integrable systems. Lie groups have a duality
between 2-form Cartan-Maurer structure equations for basis 1-forms and
Lie commutator products of dual basis vector fields.  This duality
persists when the additional cc 2-forms of an EDS are imposed,
but then the vector commutator table is incomplete.  New vectors can
be introduced in terms of the unknown commutators, and then more
commutators calculated using the Jacobi identities. These allow adding 2-form structure equations for new dual 1-forms in higher dimensional spaces.  If this expansion terminates, an embedding in a group has been
found, the new 1-forms being potentials that integrate the original
EDS. If the expansion continues, it leads to a Ka\^c-Moody
algebra of finite growth.  Such EDS belong to so-called integrable
systems of partial differential equations. The prototype of this is
the well-known Korteweg-de Vries equation, which both leads to
\cite{est90}, and belongs to, the hierarchy of the infinite Lie
algebra \({{{A_1}}^{(1)}}\)derived from SL(2, R).  The Ka\^c-Moody
algebras dual to our embedding EDS remain to be worked out.

Finally, although we did not derive these EDS variationally, Cartan forms are
easily found, at least for even dimensions.  In particular, in the (3,
7) theory either the 2-forms \({{\tau }_{A }} = {{\omega
  }_{{Ai}}}\wedge {{\theta }_i}\) or \({{\sigma }_i} = {{\omega
  }_{{iA}}}\wedge {{\theta }_{A }}\) can be used to write a quadratic
Cartan form as in some Yang-Mills theories:
\begin{equation}
\Lambda  = {{\tau }_A}\wedge {{\tau }_A},
\hspace{0.05cm}\mbox{so}\hspace{0.2cm} d{\Lambda } =
2 {{\tau }_A}\wedge {{\omega }_{{Ai}}}\wedge {{\sigma }_i}
\end{equation}
Every term of $d\Lambda$ contains both a ${{\tau }_A}$ and a ${{\sigma
  }_i}$ so arbitrary variation yields the EDS.  We also note that
${{\tau }_A}\wedge {{\tau }_A} + {{\sigma }_i}\wedge {{\sigma }_i}$ is
exact.

\section{The Partition (1, 2)}

We will set this EDS on the frame bundle ISO(1, 2) over a flat 3-space
with signature (-, \(+\), -), so the structure equations of the bases
are
\begin{eqnarray}
{d\theta }_{1} + {\omega }_{12} \wedge {\theta }_{2} + {\omega }_{31}
\wedge {\theta }_{3} &=& 0  \\
{d\theta }_{2} + {\omega }_{12} \wedge {\theta }_{1} - {\omega }_{23}
\wedge {\theta }_{3} &=& 0  \\
{d\theta }_{3} - {\omega }_{31} \wedge {\theta }_{1} - {\omega }_{23}
\wedge {\theta }_{2} &=& 0  \\
{d\omega }_{12} - {\omega }_{31} \wedge {\omega }_{23} &=& 0  \\
{d\omega }_{23} - {\omega }_{12} \wedge {\omega }_{31} &=& 0  \\
{d\omega }_{31} + {\omega }_{23} \wedge {\omega }_{12} &=& 0,
\end{eqnarray}
and the EDS to be integrated is generated by the three 2-forms
${\omega }_{iA} \wedge {\theta }^{A}, {\omega }_{Ai} \wedge {\theta
}^{i}, i = 1, A = 2, 3$:
\begin{eqnarray}
& & {\omega }_{12} \wedge {\theta }_{2} + {\omega }_{31} \wedge {\theta
}_{3}  \\
& & {\omega }_{12} \wedge {\theta }_{1}  \\
& &{\omega }_{31} \wedge {\theta }_{1}.
\end{eqnarray}

The characteristic integers are 6\{0, 3\} q \(=\)2 and CC
= 1; O(2) fiber (since ${\omega }_{23}$ is not present). To introduce
coordinates - scalar fields - we will successively prolong the EDS
with potentials or pseudopotentials, checking at each step that it
remains well-set and causal.

First, it is obvious that there is a conservation law, a closed 2-form
that is zero mod the EDS, viz. \({{{d\theta }}_1}\). So we adjoin the
1-form
\begin{equation}
{{\theta }_1}+{dv},
\end{equation}
introducing the scalar potential $v$. The characters are now $7\{1,
3\}2+1$. Next we specialize to a particular,
convenient, fiber cross-section making a choice of frame: we
introduce two new fields \(\zeta \) and \(\eta \) while prolonging
with three 1-forms taken so that the original 2-forms in the EDS
vanish (they have been ``factored'')
\begin{eqnarray}
{{\omega }_{12}}-\zeta  {{\theta }_1}  \\
{{\omega }_{13}}-\eta  {{\theta }_1}\zeta   \\
\zeta  {{\theta }_2}-\eta  {{\theta }_3}+(\eta +\zeta ){{\theta }_1}.
\end{eqnarray}

To maintain closure, however, three new 2-forms, exterior derivatives
of these or algebraically equivalent, must also be adjoined:
\begin{eqnarray}
& &(d{\zeta } - \eta {\omega }_{23}) \wedge {dv}  \\
& & (d{\eta } - \zeta {\omega }_{23}) \wedge {dv}  \\
& & (\eta {d\zeta }- \zeta {d\eta }) \wedge ({{\theta }_2}+{{\theta }_3})-
(\eta + \zeta ){\omega }_{23} \wedge (\eta {\theta }_{2} - \zeta
{\theta }_{3}).
\end{eqnarray}

Now we have $9\{4, 3\}2$ with no CC. ${\omega }_{23}$ now appears in the EDS, but is conserved, ${d\omega
}_{23}= 0$ mod EDS. Thus, we can introduce a pseudopotential variable
$x$ , and then further find another conserved 1-form and a final
pseudopotential $u$. Which is to say we can adjoin
\begin{eqnarray}
{{\omega }_{23}}-{dx}  \\
{{\theta }_2}+{{\theta }_3}-{e^x}{du},
\end{eqnarray}
without adding any 2-forms to the EDS. We have a total of 11 basis
1-forms: six in ${\theta }_{i}$, ${\theta }_{A}$, ${\omega }_{AB}$,
${\omega }_{iA}$, plus $d\zeta, d\eta, dx, du, dv$, and an EDS with
$11\{6, 3\}2$. The pulled-back original six bases are now all
solvable in terms of coordinate fields on the solutions, and can be
eliminated: $5\{0, 3\}2$. We have eliminated the CC.  This is equivalent to a set of first
order partial differential equations in 3 dependent variables and 2
independent variables.  From the character table, we expect solutions to involve 3 arbitrary functions of 1 variable.

Taking $x$ and $v$ as independent in the solution, we can solve the
first two 2-forms in Eq. (25) and (26) for \(\eta \) and \(\zeta \):
\begin{eqnarray}
\eta  &=& a {e^x}+b {e^{-x}}  \\
\zeta &=&a {e^x}-b {e^{-x}},
\end{eqnarray}
where $a$ and $b$ are arbitrary functions of $v$. The third 2-form then
amounts to
\begin{equation}
{e^x} = 1/2{{(b/a)}^\prime} {{\partial }_x}u \ ,
\end{equation}
\noindent
which integrates to
\begin{equation}
e^x = 1/2{B^\prime}(u-A(v)) \ .
\end{equation}
We have put $b/a = B(v)$ and prime is derivation with respect to $v$.

The three arbitrary functions of $v$, $a$, $b$ and $A$, give the general
solution. On it the pulled-back bases of ${E^3}$ (no longer
orthonormal or independent) are
\begin{eqnarray}
{{\theta }_1} &=& - {dv}  \\
{{\theta }_2} &=& {dv} +\frac{a {e^x}+b {e^{-x}}}{2 a}{du}  \\
{{\theta }_3} &=& - {dv} +\frac{a {e^x}-b {e^{-x}}}{2 a}{du},
\end{eqnarray}
and the induced 2-metric from \({E^3}\) is
\begin{eqnarray}
g & = & - {{\theta }_1}{{\theta }_1}+{{\theta }_2}{{\theta }_2}-{{\theta
}_3}{{\theta }_3}  \\
& = & B {{{du}}^2}+{B'}(u-A){dv} {du}- {{{dv}}^2} \ .
\end{eqnarray}

This is, up to signature, the metric found classically from the
construction of geodesically ruled surfaces in \({E^3}\), cf, e.g., Eisenhardt
[10].  The surfaces are intrinsically characterized by a ``line of
striction", the locus $u-A(v) = 0$, and a ``parameter of distribution"
$2B/{B^{\prime }}$.  The geodesic rulings, on which ${{\theta }_2}$,
${{\theta }_3}$, ${{\omega }_{12}}$, and ${{\omega }_{13}}$ pull back
to vanish, are the set of lines $u =$ const.  The rigid congruence is
the set of lines on which V contracted with ${\theta }_{1}$, ${\omega
}_{12}$ and ${\omega }_{13}$ vanishes, hence $v =$ const.

\subsection*{ACKNOWLEDGEMENTS}

This work has at every point benefitted from conversations with my
long time colleague Hugo D. Wahlquist.  It follows on several lines of
research we have jointly pursued.  Wahlquist first suggested the new
non-isometric.  embedding EDS for vacuum relativity ($q = 4$).  The
Cartan characters reported were all obtained using his suite of
{\itshape Mathematica} programs for manipulation of EDS.  The author
acknowledges a visiting appointment at the Jet Propulsion Laboratory,
California Institute of Technology


\begin{thebibliography}{10}
  
\bibitem{Ivey} Ivey, T. A. and Landsberg, J. M., \textit{Cartan for
    Beginners: Differential Geometry via Moving Frames and Exterior
    Differential Systems (Graduate Studies in Mathematics vol 61)}
  (American Mathematical Society: Providence, RI), 2003. 
  
\bibitem{Wahlquist} Wahlquist, H. D., Monte Carlo calculation of
  Cartan characters in N. Breton, R. Capovilla and T.Matos, Eds. \textit{Proceedings of the International
    Conference of Aspects of General Relativity and Mathematical
    Physics} (
  Ediciones CINVESTAV: Mexico DF), 168, 1994. 

  
\bibitem{got} Gotay, M. J., An exterior differential system approach
  to the Cartan form, symplectic geometry and mathematical physics, in P. Donato, J. Elhadad and G. M.
  Tuynmen, Eds., {\itshape Proc. Colloq. in Honor of J-M Souriau (Aix-en-Provence,
    1990) (Progress in Mathematics)} (Birkha\"{u}ser: Boston MA), 99, 1991.

  
\bibitem{GIMMsy} Gotay, M. J., Isenberg, J., Marsden, J. and Montgomery, R.,
   Momentum maps and classical relativistic fields, Part I:
  Covariant field theory, arXiv:physics/980901, 1998.
  
\bibitem{Bryant}Bryant, R., Griffiths, P. and Grossman, D., \textit{Exterior Differential Systems and Euler-Lagrange Partial Differential Equations} (The University of Chicago Press: Chicago), 2003.
  
\bibitem{her88} Hermann, R., Differential form methods in the theory
  of variational systems and Lagrangian field theories, {\itshape Acta
    Appl. Math.}{\bfseries {\itshape 12}}, 35-78, 1988.
  
\bibitem{est80} Estabrook, F. B., Differential geometry as a tool for
  applied mathematics, in R. Martini, Ed. {\itshape Lecture Notes in Math. }{\bfseries
    {\itshape 810}} (Springer: Berlin), 1980.
      
\bibitem{est06} Estabrook, F. B., Conservation laws for vacuum tetrad
  gravity, \textit{Class. Quantum Grav.} {\bfseries
    {\itshape 114}}, 2841-2848,  2006.
  
\bibitem{bry83} Bryant, R., Griffiths, P. A.. and Yang, D.,
  Characteristics and existence of isometric embeddings, {\itshape Duke Math. J.} {\bfseries {\itshape 50}}, 893-994, 1983.
  
\bibitem{gri87} Griffiths, P. A. and Jensen, G. R., Differential systems
  and isometric embeddings, {\itshape Annals of Mathematical Studies} {\bfseries
    {\itshape 114}} (Princeton University Press: Princeton NJ), 1987.
  
\bibitem{est89} Estabrook, F. B. and Wahlquist, H. D., Classical
  geometries defined by exterior differential systems on higher frame
  bundles, {\itshape Class. Quantum Grav. }{\bfseries {\itshape 6
    }}, 263-274, 1989.
  
\bibitem{est99} Estabrook, F. B., Robinson, R. S. and Wahlquist, H. D., Constraint-free theories of gravitation, {\itshape Class. Quantum
    Grav. }{\bfseries {\itshape 16 }}, 911-918, 1999.
  
\bibitem{est05} Estabrook, F. B., Mathematical structure of tetrad
  equations for vacuum relativity, {\itshape Physical Review}
  {\bfseries
    {\itshape D 71}}, 044004, 2005.
  
\bibitem{est82} Estabrook, F. B., Moving frames and prolongation
  algebras, {\itshape Journal of Mathematical Physics} {\bfseries
    {\itshape 23}}, 2071-76, 1982.
 
\bibitem{har83} Harrison, B. K., Ernst equation Backlund
  transformations revisited: New approaches and results, in Hu Ning,
  Ed. {\itshape Proceedings of the Third Marcel Grossman Meeting on
    General Relativity} (North-Holland, Amsterdam), 29-39, 1983. 
  
\bibitem{har84} Harrison, B. K., Prolongation structures and
  differential forms, in C. Hoenselaers and W. Dietz, Eds.,{\itshape Solutions of Einstein's Equations:
  Techniques and Results, Lecture Notes in Physics} \#205  (Springer, Berlin), 26-54, 1984.
  
\bibitem{wah66} Wahlquist, H. D. and Estabrook, F. B., Rigid motions in
  Einstein space, {\itshape J. Math. Phys.}{\bfseries {\itshape 7,}} 894-905, 1966.
  
\bibitem{est90} Estabrook, F. B., Differential geometry techniques
  for sets of nonlinear partial differential equations, in R. Conte and
  N. Boccara, Eds., {\itshape Partially Integrable Evolution Equations
    in Physics} (Kluwer: Dordrecht), 413-434, 1990.
  
\bibitem{eis60} Eisenhart, L. P., {\itshape A Treatise on the
    Differential Geometry of Curves and Surfaces} (Dover: New York), 241 {\itshape et
    seq}, 1909, 1960.  
\end{thebibliography}
\end{document}